\documentclass[10pt]{article} 
\usepackage{amsmath,amsfonts,amssymb,slashed,makeidx,latexsym,setspace}
\usepackage{graphicx,graphics,floatflt,amssymb,epsf,rotate,subfigure} 
\usepackage{times,axodraw,cite,mathrsfs,wrapfig}
\evensidemargin=\oddsidemargin 
\usepackage{color}
\def\L{\hat{L}}

\def\O{\cal{O}}

\def\g{\gamma}

\def\L{\Lambda}

\begin{document} 
\begin{flushright} 
SINP/TNP/2012/02
\end{flushright} 
 
%\vskip 20pt 
 
\begin{center}
  {\Large \bf Electroweak Symmetry Breaking Beyond the Standard Model
    \footnote{Plenary talk at the International
      Lepton-Photon Conference, Mumbai, August 2011.  To appear
      in the proceedings (special issue of PRAMANA).} } \\
  \vspace*{1cm} \renewcommand{\thefootnote}{\fnsymbol{footnote}}
  {\bf Gautam Bhattacharyya} \\
  \vspace{10pt} {\small {\em Saha Institute of Nuclear Physics, 1/AF
      Bidhan Nagar, Kolkata 700064, India}}
    
\normalsize 
\end{center} 

\begin{abstract}  
  
  In this talk, I shall address two key issues related to electroweak
  symmetry breaking. First, how fine-tuned different models are that
  trigger this phenomenon?  Second, even if a light Higgs boson
  exists, does it have to be necessarily elementary? After a brief
  introduction, I shall first review the fine-tuning aspects of the
  MSSM, NMSSM, generalized NMSSM and GMSB scenarios. I shall then
  compare and contrast the little Higgs, composite Higgs and the
  Higgsless models. Finally, I shall summarize by giving a broad
  overview on where we stand at the end of 2011.

%\vskip 5pt \noindent 
%\texttt{PACS Nos:~ 12.60.Jv, 11.10.Kk } \\ 
%\texttt{Key Words:~~Higgs, Supersymmetry, Extra Dimension}
\end{abstract}

\renewcommand{\thesection}{\Roman{section}} 
\setcounter{footnote}{0} 
\renewcommand{\thefootnote}{\arabic{footnote}}

\section{Introduction}
\begin{wrapfigure}[12]{r}{0.5\textwidth} 
\epsfxsize=0.45\textwidth
\centering \rotatebox{0}{\epsfbox{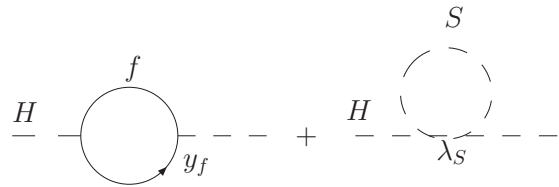}}
  \caption{\small \em Cancellation of quadratic divergence to scalar
    mass-square between fermion and boson loops.}
\label{cancel} 
\end{wrapfigure}%

The timing of the last `Lepton-Photon Conference' (August 2011) was
very special! Every day the LHC was delivering more data than it did
during the entire 2010. The time for `speculation' was soon coming to
an end!  Our imagination about the possible dynamics behind
electroweak symmetry breaking (EWSB), disciplined by the constraints
from electroweak precision tests (EWPT), has fueled different
directions of theoretical studies and experimental searches over the
last so many years. Finally, the LHC has roared into life, and this is
our last chance of putting money on our favorite models. It is in
this backdrop that I have prepared a write-up of my talk, being aware
that even during the last few months since Lepton-Photon the excluded
territory for different Beyond the Standard Model (BSM) alternatives
has further grown in size.

Now, to the point. We know that the SM Higgs mechanism is only an
effective description of EWSB.  Can LHC shed enough light on the
dynamics behind this mechanism?  Some of the questions that drive our
speculation are listed below \cite{arXiv:0910.5095}:

\noindent $(i)$~ Why is the weak scale so much separated from the
Planck scale?

\noindent $(ii)$~ What is the symmetry that controls particle physics
at the TeV scale? In other words, now that the gauge symmetry is
established with a significant precision, what is the next relevant
symmetry that awaits us?

\noindent $(iii)$~ The SM is plagued by the hierarchy problem. It
originates from the requirement of {\em ad hoc} cancellation between
fermionic and bosonic loops contributing to the Higgs mass -- see
Fig.~\ref{cancel}.  An unnatural tuning ($1 \div 10^{26}$) between
the bare Higgs mass-square $m_{h_0}^2$ and the correction term $\Delta
m_h^2$ is necessary to keep the renormalized mass ($m_h^2 = m_{h_0}^2
+ \Delta m_h^2$) at around 100 GeV. Nevertheless, one must do this
tuning order-by-order in perturbation theory to prevent the Higgs mass
from shooting up to the highest scale of the theory.  This constitutes
the hierarchy problem. Quite a few remedies have been advocated so
far. But, which solution (if any, at all!) of the hierarchy problem is
correct?

\noindent $(iv)$~ Is the naturalness consideration a good guiding
principle or a powerful discriminator between models? Is its study a
step in the right direction \cite{Giudice:2008bi}?

\noindent $(v)$~ Is Higgs {\em elementary} or {\em composite}
\cite{Contino:2010rs,arXiv:0910.4976}? Can it be settled at the LHC?

\noindent $(vi)$~ What if the Higgs is not there at all?

\section{Supersymmetry}
\subsection{Basic aspects}
Supersymmetry is the most well-studied BSM model that offers a natural
explanation of the weak scale \cite{susy-books}. It is a new
space-time symmetry interchanging bosons and fermions, relating states
of different spins.  The Poincare group is extended by adding two
anti-commuting generators $Q$ and $\bar{Q}$ to the existing $p$ (linear
momentum), $J$ (angular momentum) and $K$ (boost), such that
$\{Q,\bar{Q}\} \sim p$. Since the new symmetry generators are spinors,
not scalars, supersymmetry is not an internal symmetry, and the
super-partners differ from their SM partners in spin. Some attractive
features of supersymmetry relevant in the present context are as
follows:

\noindent $(i)$~ {\em Supersymmetry solves the gauge hierarchy
  problem}: The quantum corrections to the Higgs mass from a bosonic
loop and a fermionic loop exactly cancel if the couplings are
identical and the boson is mass degenerate with the fermion. For every
fermion (boson) of the SM, spersymmetry provides a mass degenerate
boson (fermion).  In real life, however, supersymmetry is badly
broken. But if the breaking occurs in masses and not in dimensionless
couplings, the quadratic divergence still cancels. The residual
divergence is only logarithmically sensitive to the supersymmetry
breaking scale.

\noindent $(ii)$~{\em Supersymmetry leads to gauge coupling
  unification}: This is a bonus! Supersymmetry was not invented to
achieve this.  When the SM gauge couplings are extrapolated to high
scale, with LEP measurements as input, they do not meet at a single
point. Supersymmetry makes them do at a scale $M_{\rm GUT} \sim 2
\times 10^{16}$ GeV, with TeV scale super-particles. 

\noindent $(iii)$~ {\em Supersymmetry triggers EWSB}: Starting from a
positive value in the ultraviolet, the up-type Higgs mass-square
$m_{H_u}^2$ turns negative in the infrared triggering EWSB. In the SM
the negative sign in front of the scalar mass-square in the potential
is put in by hand to ensure EWSB. In supersymmetry the sign flip
occurs in a dynamical way.

\subsection{Naturalness criterion}
Naturalness is an aesthetic criterion. It comes from the realization
that if large cancellation among {\em unrelated} quantities is
required to achieve a small physical quantity, the situation is
unnatural and reflects a sign of {\em weak health} of the theory.  A
theory is less `natural' if it is more `fine-tuned'. In the context of
minimal supersymmetry with two Higgs doublet (i.e. MSSM), the scalar
potential minimization yields
\begin{eqnarray}
\frac{1}{2} M_Z^2 = \frac{m_{H_d}^2 - m_{H_u}^2 \tan^2\beta}
{\tan^2\beta - 1} - \mu^2,
\label{nat}
\end{eqnarray}
 with $m_{H_u}^2 = m_{H_d}^2 - \Delta m^2$, where $\Delta m^2$ is the
 correction due to RG running from the GUT scale to the weak
 scale. The large top Yukawa coupling has a significant numerical
 influence on RG running.  A proper EWSB occurs when $m_{H_u}^2$ turns
 negative due to the effect of running and the correct value of $M_Z$
 is reproduced. This refers to a cancellation between supersymmetry
 breaking soft masses and supersymmetry preserving $\mu$ parameter.
 How much cancellation between these completely uncorrelated
 quantities is aesthetically pleasant?  Barbieri and Giudice
 introduced a quantitative measure of fine-tuning
 \cite{CERN-TH-4825/87}
\begin{eqnarray}
\Delta_i \equiv
\left|\frac{\partial M_Z^2/M_Z^2}{\partial a_i/a_i}\right|,
\label{delta}
\end{eqnarray}
where $a_i$ are high scale input parameters.  An upper limit on
$\Delta$ can be translated to an upper limit on super-particle masses.

\subsection{Naturalness of cMSSM}
In cMSSM, the constrained version of the MSSM (with 4
parameters and 1 sign), Eq.~(\ref{nat}) boils down to \cite{arXiv:1101.2195}
\begin{eqnarray}
M_Z^2 \approx  -2 |\mu^2| + 0.2~m_0^2 + 0.7 \left(2.6~ M_{1/2}\right)^2, 
\label{cmssmnat}
\end{eqnarray}
where  $m_0$ and
$M_{1/2}$ are the common scalar and gaugino masses, respectively, 
and the gluino mass is given by $m_{\tilde{g}} \simeq
2.6~M_{1/2}$. Two observations are noteworthy:

\noindent $(i)$~ In the absence of any cancellation, the natural
expectation would be $M_Z \sim \mu \sim m_0 \sim M_{1/2}$. But this
possibility has been explored and ruled out by LEP-2 and Tevatron.

\noindent $(ii)$~ By now, the CMS and ATLAS collaborations have pushed
the gluino mass limit to close to a TeV. This implies a tuning of
order 1\% from Eq.~(\ref{cmssmnat}). The LHC is thus probing sparticle
masses which are about a loop factor above $M_Z$.

There is another way to show that the fine-tuning in MSSM is $\sim
1\%$ level.  The radiatively corrected mass of the lightest CP-even
Higgs boson is given by
\begin{equation}
m_h^2 \simeq m_{h0}^2 (\leq M_Z^2) + \frac{3 m_t^4}{2 \pi^2 v^2} 
\ln\left( \frac{m^2_{\tilde{t}}}{m_t^2}\right).
\end{equation}
Since $m_h > 114$ GeV from LEP-2, $m_{\tilde{t}}$ should be around 1
TeV or heavier, thus implying a fine-tuning to the tune of a
per-cent. This constitutes the `little hierarchy' problem of
supersymmetry.

%%%%%%%%%%%%%%%%%%%%%%%%%%%%%%%%%%%%%%%%%%%%%%%%%%%%%%%%%%%%%%%%%%%%
\begin{figure*}
%\vspace{-0.1in}
\begin{minipage}[t]{0.42\textwidth}
\epsfxsize=6cm
\centering{\epsfbox
{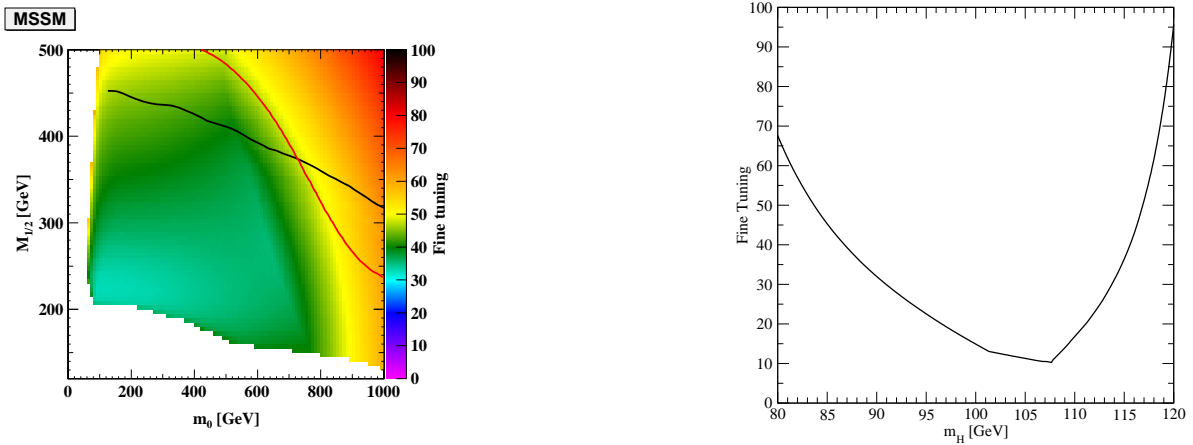}}
\end{minipage}
\hfill
\begin{minipage}[t]{0.42\textwidth}
\epsfxsize=6cm
\centering{\epsfbox
{ellwanger_fig2.eps}}
\end{minipage}
\vspace{5pt}
\caption[]{\small{\em (a) Left: Fine-tuning in cMSSM. (b) Right:
    Fine-tuning as a function of $m_h$. For details see
    \cite{arXiv:1107.2472}.}}
\label{ellwanger}
\end{figure*}
%%%%%%%%%%%%%%%%%%%%%%%%%%%%%%%%%%%%%%%%%%%%%%%%%%%%%%%%%%%%%%%%%%%%%%%%%

A quantitative analysis of fine-tuning has recently been carried out
in \cite{arXiv:1107.2472} (see also \cite{arXiv:1101.4664} where some
of the technical aspects for measuring the tuning are a little
different) in the context of the cMSSM. Fig.~\ref{ellwanger}a
corresponds to $\tan\beta = 3$ and $A_0 = 0$. The different parts of
the white region is ruled out for different reasons (non-occurrence of
EWSB, experimental exclusion of the slepton/neutralino/chargino mass
limits, Higgs mass lower limit, stau becoming the LSP). The
experimental bounds from ATLAS (black) and CMS (red) have been drawn
for a guide to the eye using $1/fb$ data. The fine-tuning is at best
$\sim$ 2\% which corresponds to $\Delta \sim 50$.  We now look at
Fig.~\ref{ellwanger}b where fine-tuning has been plotted against
$m_h$.  The LEP-2 lower limit has not been imposed here. It is
interesting to see that the tuning is minimum around $m_h = 108$
GeV. If $m_h$ is lower than that, the fine-tuning becomes larger as
sparticle masses are constrained by their experimental lower
limits. On the other hand, if $m_h$ is higher than this value, then
due to its $\ln(m_{\tilde t}$-dependence there is an exponential
growth of fine-tuning.

It is interesting to note that for values of $m_0 \leq 700$ GeV and
$M_{1/2} \leq 350$ GeV, the amount of fine-tuning is decided by the
LEP-2 limit on the Higgs mass. On the other hand, for larger $m_0$ and
$M_{1/2}$, the origin of fine-tuning can be traced to the adjustment
between $\mu^2$ and scalar soft mass-squares that yields the correct
$M_Z$.

\subsection{Naturalness of NMSSM}
First we consider the NMSSM scenario which has an additional gauge
singlet superfield $S$ compared to MSSM \cite{nmssm}. The NMSSM
superpotential has two important additional pieces,
\begin{eqnarray} 
W_{\rm NMSSM} = W_{\rm Yukawa} + \lambda S H_u H_d + \frac{1}{3}
\kappa S^3 \, .
\label{nmssm-w}
\end{eqnarray}  
The vev $s$ of the scalar component of $S$ yields an effective
$\mu_{\rm eff} = \lambda s$. In fact, this was the main motivation
behind adding the singlet. The NMSSM models are less fine-tuned than
MSSM for three reasons \cite{arXiv:1107.2472}:

\noindent $(i)$ The $SH_u H_d$ term in Eq.~(\ref{nmssm-w}) generates a
quartic interaction in the scalar potential, increasing tree level
$m_{h_0}$ \cite{Drees:1988fc},
\begin{eqnarray}
m_{h0}^2 \approx M_Z^2 \cos^2 2\beta + \lambda^2 v^2 \sin^2 2\beta \, .
\label{nmssm-mass}
\end{eqnarray} 
The additional tree level contribution allows us to consider a lighter
stop in the loop to generate the same Higgs mass as in the MSSM.
Fine-tuning is therefore reduced.

\noindent $(ii)$ The physical Higgs boson can have a large singlet
admixture, and therefore, a reduced gauge coupling which helps it
evade the LEP-2 limit.  This again implies that we can employ a
lighter stop in the loop, thus reducing fine-tuning.

\noindent $(iii)$ The possibility of Higgs decaying into two lighter
pseudo-scalars also helps to evade the LEP-2 limit.

\begin{figure*}
%\vspace{-0.1in}
\begin{minipage}[t]{0.45\textwidth}
\epsfxsize=6cm
\centering{\epsfbox
{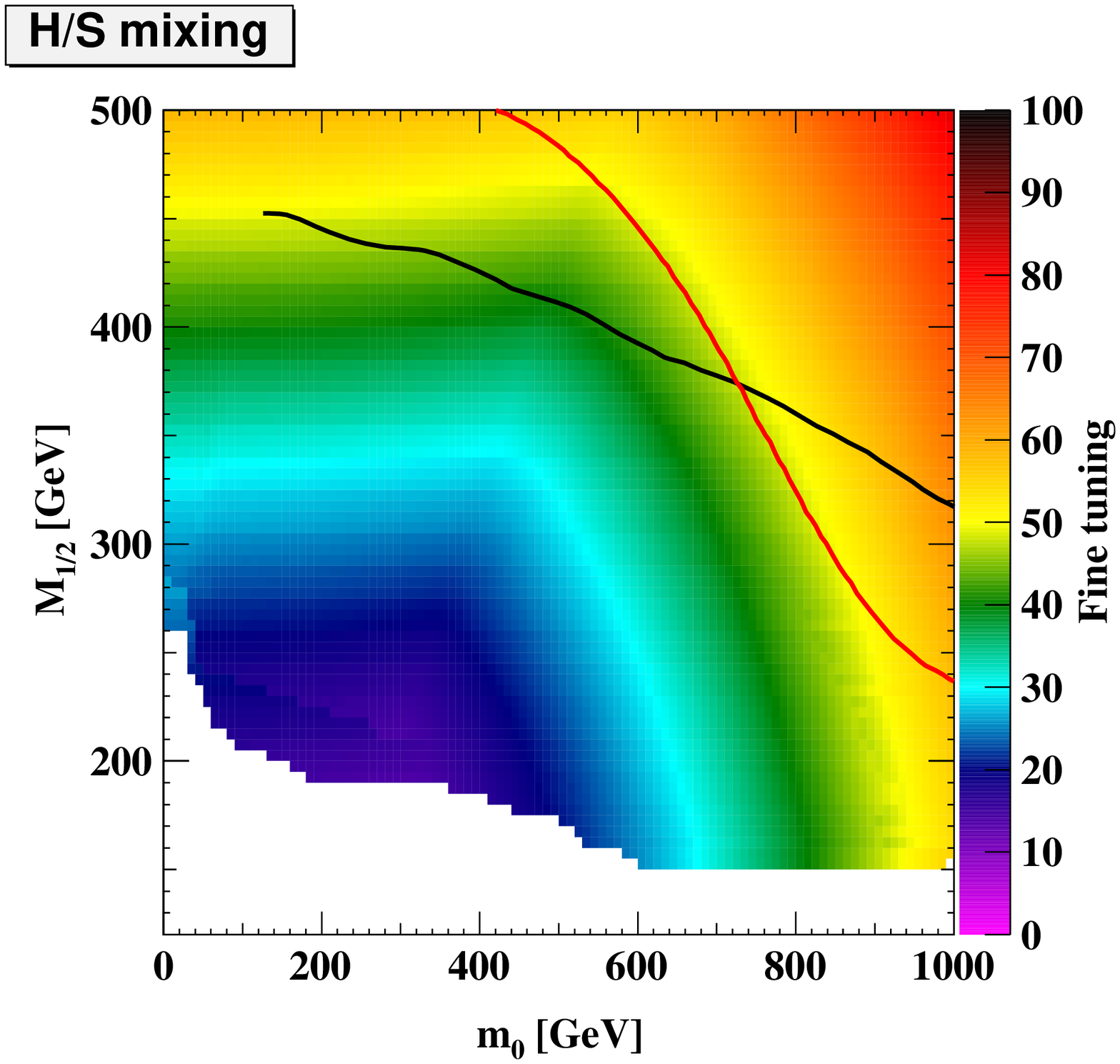}}
\end{minipage}
\hfill
\begin{minipage}[t]{0.45\textwidth}
\epsfxsize=7cm
\centering{\epsfbox
{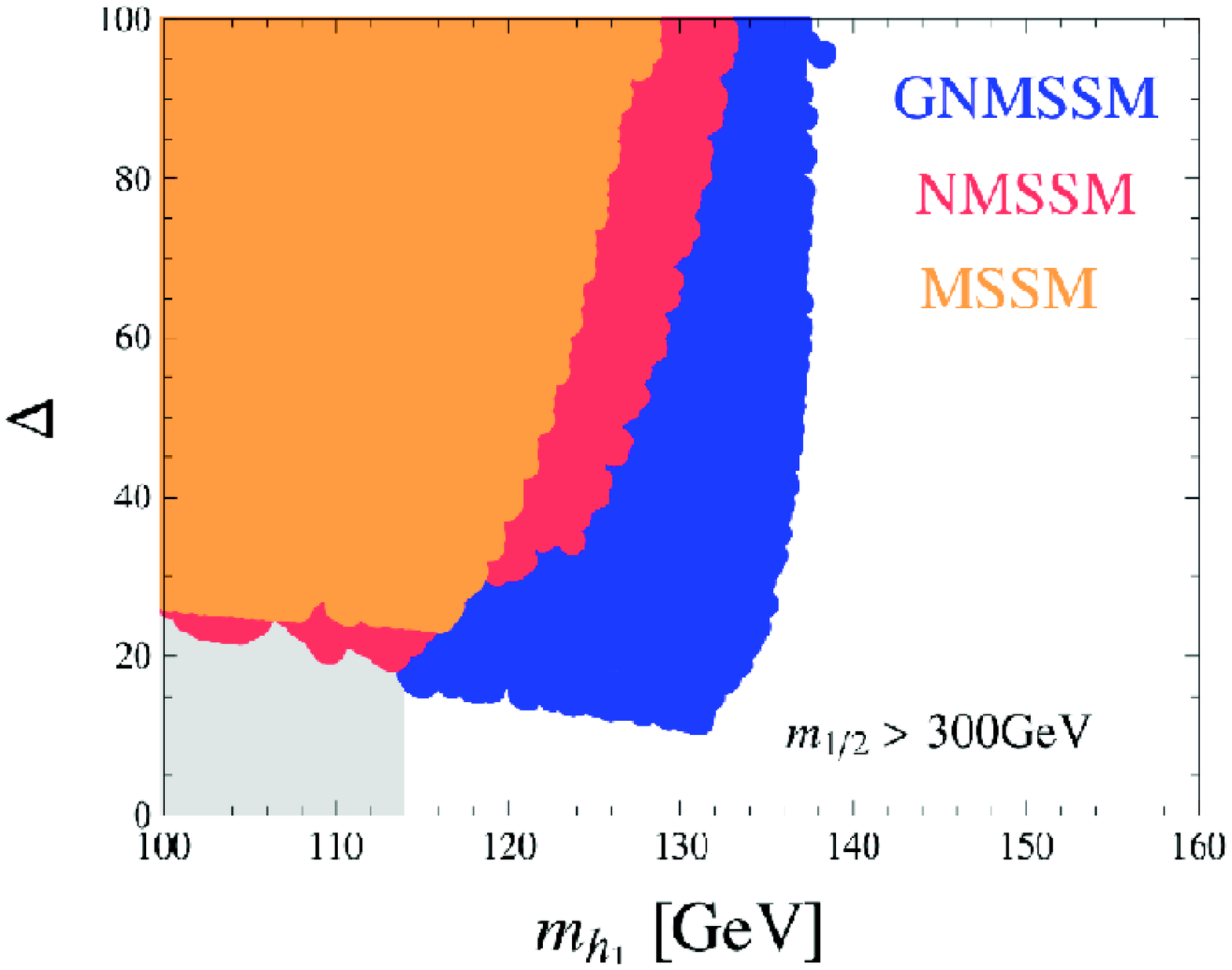}}
\end{minipage}
\vspace{5pt}
\caption[]{\small{\em (a) Left: Fine-tuning in NMSSM
    \cite{arXiv:1107.2472}. (b) Right: Fine-tuning in G-NMSSM
    \cite{arXiv:1108.1284}.}}
\label{gnmssm}
\end{figure*}

The minimal fine-tuning in NMSSM has been plotted in
Fig.~\ref{gnmssm}a. For smaller values of $m_0$ and $M_{1/2}$, i.e. in
the region where the LEP-2 limit on $m_h$ is the relevant constraint,
fine-tuning is considerably less than in cMSSM.  $\Delta$ can be as
small as 14 in this region (as against 33 for cMSSM). However, for
larger values of $m_0$ and $M_{1/2}$ the origin of fine-tuning lies in
the smallness of weak scale compared to the soft masses, and in this
region it is hard to reduce fine-tuning. Overall, NMSSM is less
fine-tuned than cMSSM, or for that matter in MSSM with universal
boundary conditions.

%%%%%%%%%%%%%%%%%%%%%%%%%%%%%%%%%%%%%%%%%%%%%%%%%%%%%%%%%%%%%%%%%%%%
%%%%%%%%%%%%%%%%%%%%%%%%%%%%%%%%%%%%%%%%%%%%%%%%%%%%%%%%%%%%%%%%%%%%%%%%%

\subsection{Naturalness of Generalized NMSSM (G-NMSSM)}
G-NMSSM has an underlying $Z_4$ or $Z_8$ discrete symmetry
\cite{arXiv:1108.1284}. Its superpotential reads
\begin{eqnarray}
W_{\rm G-NMSSM} &=& W_{\rm Yukawa} + (\mu + \lambda S) H_u H_d +
\frac{1}{2} \mu_S S^2 + \frac{1}{3} \kappa S^3 \, , ~~{\rm where} ~~
\mu \sim \mu_S \sim {\cal{O}} (m_{3/2}) \, .
\end{eqnarray}

It has two distinct advantages beyond NMSSM. First, it has a discrete
$R$ symmetry, contrary to a discrete -- but non-$R$ -- symmetry in
NMSSM, which helps to remove the domain wall problems present in NMSSM
\cite{hep-ph/9506359}. The $R$ symmetry is broken at a very high scale
making the domain walls decay well before nucleosynthesis. And
secondly, fine-tuning in G-NMSSM is considerably less than in
NMSSM. The main reason behind this is the additional stabilizing terms
in the potential. To appreciate this, take a large $\mu_S$ limit and
integrate out the $S$ superfield at the supersymmetric level. This
gives a term $\lambda^2 (H_u H_d)^2/\mu_S$ in the superpotential,
which reduces fine-tuning.  For a fixed value of $\Delta$ one gets a
heavier Higgs and, interestingly enough, the fine-tuning is minimum
for $m_h \sim 130$ GeV. Fig.~\ref{gnmssm}b shows us how fine-tuning
improves from MSSM to NMSSM and from NMSSM to G-NMSSM.

\subsection{Naturalness of $\lambda$SUSY}
\begin{figure*}
%\vspace{-0.1in}
\begin{minipage}[t]{0.45\textwidth}
\epsfxsize=5cm
\centering{\epsfbox
{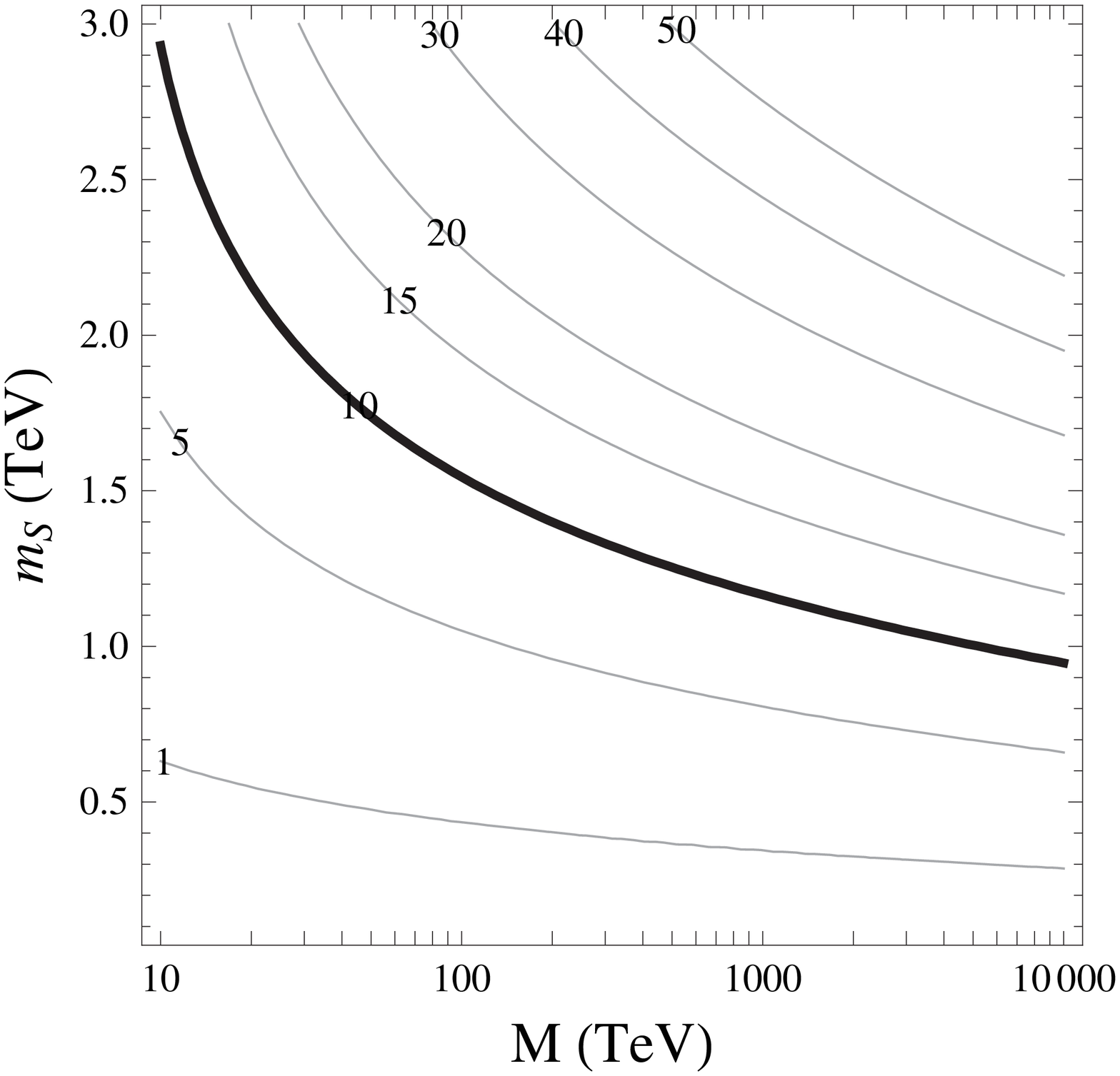}}
\end{minipage}
\hfill
\begin{minipage}[t]{0.45\textwidth}
\epsfxsize=6cm
\centering{\epsfbox
{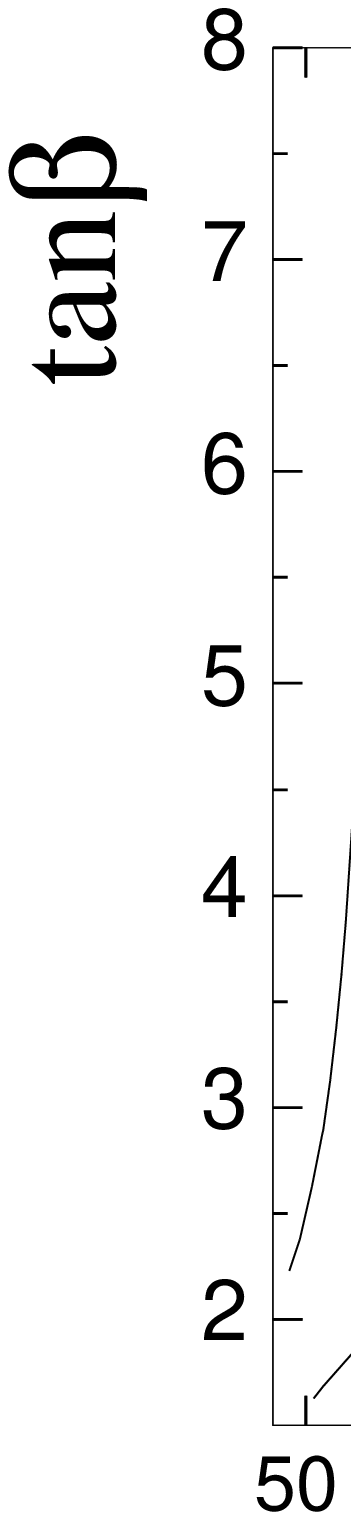}}
\end{minipage}
\vspace{5pt}
\caption[]{\small{\em (a) Left: Fine-tuning in $\lambda$SUSY
    \cite{arXiv:1004.1271}. (b) Right: Fine-tuning in GMSB
    \cite{hep-ph/9611243}.}}
\label{lambdasusy_gmsb}
\end{figure*}

Consider the NMSSM and assume that the trilinear coupling $\lambda$ is
rather large \cite{arXiv:1004.1271}, at least $\sim 1$ at the weak
scale. The sole purpose here is to reduce fine-tuning by increasing
the singlet-induced tree level contribution to the Higgs mass, so that
the dominant term is $m_{h0}^2 \sim \lambda^2 v^2 \sin^2 2\beta$. For
example, the values $m_h^{\rm max} \simeq 2(3) M_Z$ for $\lambda (\L)
= \sqrt{4\pi}$ correspond to $\Lambda = 10^4$ TeV (100 TeV) -- see
Fig.~\ref{lambdasusy_gmsb}a. The flip side is that by having such a low
cutoff, the prized possession of supersymmetry, namely, gauge coupling
unification, is sacrificed to buy naturalness!

%%%%%%%%%%%%%%%%%%%%%%%%%%%%%%%%%%%%%%%%%%%%%%%%%%%%%%%%%%%%%%%%%%%%
%%%%%%%%%%%%%%%%%%%%%%%%%%%%%%%%%%%%%%%%%%%%%%%%%%%%%%%%%%%%%%%%%%%%%%%%%

\subsection{Naturalness of GMSB models} 
In gauge mediated supersymmetry breaking (GMSB) models, a natural
determination of $M_Z$ in terms of the model parameters yields a
rather upper limit on the mass of the right selectron -- see
Fig.~\ref{lambdasusy_gmsb}b.  The universal boundary conditions of
scalar masses in cMSSM do not permit the lightest scalar to be much
lighter than $m_{H_u}^2$. But in gauge mediated models, the
proportionality of scalar masses to different gauge couplings (square)
at the messenger scale $M$ creates quite a bit of splitting among the
different scalar masses at the weak scale, which in turn leads to {\em
  more} fine-tuning than in cMSSM \cite{hep-ph/9611243}.

\section{Little Higgs}
\subsection{Basic aspects}
Little Higgs models were introduced as a solution to the little
hierarchy problem (for a review, see
\cite{lhreviews,Cheng:2007bu}). The Higgs is considered to be a
pseudo-Goldstone boson associated with some global symmetry breaking.
A Goldstone boson $\phi$ has a shift symmetry $\phi \to \phi + c$,
where $c$ is a constant, and as long as this symmetry is maintained a
Goldstone remains massless at all order. But if there is an
interaction which couples $\phi$ not as $\partial_\mu \phi$ the shift
symmetry is explicitly broken and the Goldstone becomes massive. This
way we get a pseudo-Goldstone boson.  Recall that pion is a Goldstone
which results from the spontaneous breaking of chiral symmetry group
${\rm SU(2)}_L \times {\rm SU(2)}_R$ to the isospin group ${\rm
  SU(2)}_I$.  Since quark masses and electromagnetic interaction
explicitly break the chiral symmetry, pions are in fact
pseudo-Goldstone bosons.  Electromagnetism attributes a mass to
$\pi^+$ of order $m_{\pi^+}^2 \sim (e^2 /16\pi^2)\Lambda_{\rm
  QCD}^2$. If we think of Higgs mass generation in the same way, using
gauge or Yukawa interaction as a source for explicit breaking of the
chiral symmetry, we can have $m_h^2 \sim (g^2 /16\pi^2)\Lambda_{\rm
  NP}^2$.  This picture is not phenomenologically acceptable, since
$m_h \sim 100$ GeV implies $\Lambda_{\rm NP} \sim 1$ TeV, but such a
low cutoff is strongly disfavored by EWPT. If, on the other hand, we
can somehow arrange that the leading term in the Higgs mass is
\begin{eqnarray} 
m_h^2 \sim \frac{g_1^2 g_2^2}{(16\pi^2)^2} \Lambda_{\rm NP}^2 \, ,
\label{mhlh}
\end{eqnarray} 
then for a 100 GeV Higgs mass, we get $\Lambda_{\rm NP} \sim 10$
TeV. The cutoff is thus postponed from 1 to 10 TeV thanks to the extra
suppression factor of $16 \pi^2$, without having to apparently pay any
price for fine-tuning. The idea of `little Higgs' is all about
achieving this extra $16 \pi^2$ factor in the denominator of
Eq.~(\ref{mhlh}), and this is where it differs from a pion. Note that
both $g_1$ and $g_2$ should be simultaneously non-vanishing in order to
generate the Higgs mass. If any of these couplings vanishes then the
global symmetry is partially restored and the Higgs remains a
Goldstone boson. This is the concept of `collective symmetry
breaking'.

\begin{figure*}
%\vspace{-0.1in}
\begin{minipage}[t]{0.42\textwidth}
\begin{center}
\begin{picture}(100,100)(-50,-10)
\SetWidth{1}
\CArc(0,0)(50,0,360)
\CArc(-20,0)(30,0,360)
\CArc(20,0)(30,0,360)
\Text(0,40)[]{\bf G}
\Text(-30,0)[]{\bf F}
\Text(0,0)[]{\bf I}
\Text(30,0)[]{\bf H}
    \Text(0,-71)[]{\bf (a)}
\end{picture}
\end{center}
%%%\epsfxsize=5cm
%%%\centering{\epsfbox
%%%{coset.eps}}
\end{minipage}
%\hfill
\hspace{0.5cm}
\begin{minipage}[t]{0.42\textwidth}
\begin{picture}(100,70)(-50,0)
\SetWidth{1}
    \CArc(0,40)(20,0,360)
    \DashLine(-20,40)(-40,40)2 
    \DashLine(20,40)(40,40)2 
    \Text(-30,50)[]{$h$}
    \Text(30,50)[]{$h$}
    \Text(0,70)[]{$t,T$}
\end{picture}
%\quad \quad 
\begin{picture}(100,70)(-50,0)
\SetWidth{1}
    \CArc(0,40)(20,0,360)
    \DashLine(-40,20)(40,20)2 
    \Text(-30,30)[]{$h$}
    \Text(30,30)[]{$h$}
    \Text(0,70)[]{$T$}
\end{picture}
\\
\begin{picture}(100,70)(-50,0)
\SetWidth{1}
    \PhotonArc(0,40)(20,0,360){4}{20}
    \DashLine(-40,20)(40,20)2 
    \Text(-30,30)[]{$h$}
    \Text(30,30)[]{$h$}
    \Text(0,70)[]{$\g,W,Z$}
\end{picture}
\begin{picture}(100,70)(-50,0)
\SetWidth{1}
    \PhotonArc(0,40)(20,0,360){4}{20}
    \DashLine(-40,20)(40,20)2 
    \Text(-30,30)[]{$h$}
    \Text(30,30)[]{$h$}
    \Text(0,70)[]{$A_H,W_H,Z_H$}
    \Text(-50,8)[]{\bf (b)}
\end{picture}
\end{minipage}
\vspace{5pt}
\caption[]{\small{\em (a) Left: Little Higgs cartoon.  (b) Right:
    Feynman diagrams among which same statistics cancellation
    takes place. $T$ is a new heavy quark, and $A_H, W_H, Z_H$ are new
    heavy gauge bosons.}}
\label{lhfig}
\end{figure*}
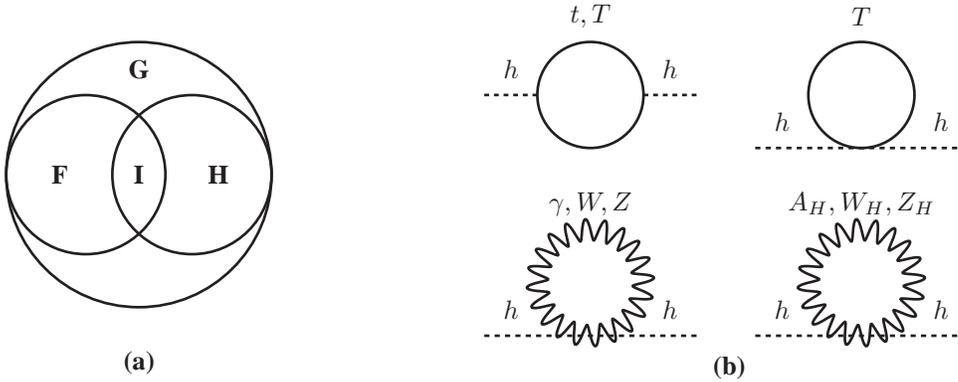

The basic features of the little Higgs trick are depicted in
Fig.~\ref{lhfig}a.  The global group $G$ spontaneously breaks to $H$
at a scale $f (>v)$.  A part of $G$, labeled $F$, is weakly gauged
and the overlap region between $F$ and $H$ is the unbroken SM group
$I$. The Higgs, which is a doublet of the gauged SU(2) of the SM, is a
part of the Goldstone multiplet that parametrizes the coset space
$G/H$.  The generators corresponding to Higgs do not commute with the
heavy gauge boson generators. Gauge (also, Yukawa) interactions induce
mass to the Higgs boson at one-loop level.  Since the gauge group is
expanded, we have additional gauge bosons and fermions. The quadratic
divergence to the Higgs mass at one-loop level arising from a $Z$
boson loop cancels against a similar contribution from a heavy $Z_H$
loop, and the same thing happens between a $t$ loop and a heavy $T$
loop -- see Fig.~\ref{lhfig}b. This is an example of `same
  statistics cancellation'.

\subsection{Two crucial features}
\noindent $(i)$~ The same statistics cancellation enables us to
express $m_h^2 \sim f^2/16\pi^2 \ln(\L^2/f^2)$. But the quadratic
cutoff sensitivity comes back parametrically at two-loop order.  The
order parameter is $f$ is not protected from quadratic cutoff
sensitivity, just like the electroweak vev $v$ is not
\cite{Kaul:2008cv}. As a result,
    \begin{equation}
       f^2 \to F^2 = f^2 + \frac{\Lambda^2}{16\pi^2}  
~~~~~({\rm where}~ \L \sim 4\pi f) \, ,  
     \end{equation}
Then, what did we gain compared to the SM?  For little Higgs models
\begin{equation}
\label{mhlh-2}
  m_h^2~({\rm LH}) \sim \left(\frac{F^2}{16\pi^2}\right) 
 \ln \left(\frac{\L^2}{F^2}\right)   \, . 
\end{equation}
This implies that $\Delta m_h^2~({\rm LH}) \sim \L^2/(16\pi^2)^2$,
which should be compared with $m_h^2 ({\rm SM}) \sim
\L^2/(16\pi^2)$. For little Higgs, we thus have an extra suppression
factor of $16\pi^2$, which indicates the parametric two-loop
sensitivity of $\L^2$.  If we want $m_h \sim (f/4\pi) \sim 100$ GeV,
one should have $f \sim F \sim 1$ TeV, and $\Lambda \sim 10$ TeV.

\noindent $(ii)$~ A clever construction of a little Higgs model
should yield the following electroweak potential:
\begin{equation}
  V = - \frac{\left({g_{\rm SM}}\right)^4}{16\pi^2}f^2 
\ln\left(\frac{\L^2}{f^2}\right)
  (H^\dagger H) + g_{\rm SM}^2 (H^\dagger H)^2 \, , 
\end{equation} 
i.e., the bilinear term should have a one-loop suppression but,
crucially, the quartic interaction should be {\em un-suppressed},
where $g_{\rm SM}$ is a gauge or Yukawa coupling. If both quadratic
and quartic terms are suppressed, one cannot simultaneously obtain the
correct $W$ boson mass and an acceptable Higgs mass.

\subsection{EWPT vs Naturalness}  
Contributions of new physics to two dimension-6 operators ${\O}_T
\propto \left|H^\dagger D_\mu H \right|^2$ and ${\O}_S \propto
H^\dagger \sigma^a H W_{\mu\nu}^a B_{\mu\nu}$ should be small enough
to keep EWPT ($T$ and $S$ parameters, respectively) under control. A
large class of little Higgs models gives a large contribution to
$T$. Consequently, the constraint is quite strong: $f > (2-5)$ TeV
\cite{large_f_ewpt}. A large $f$ means that to obtain the Higgs mass
in the 100 GeV range one must fine-tune the parameters. The
constraints arise primarily from the tree level mixing of the SM
particles with the new particles. In the littlest Higgs model ($G =
{\rm SU(5)}, ~ H = {\rm SO(5)}$), the $T$ parameter receives a large
contribution from the custodial symmetry breaking operator $H^T \Phi
H$, which mixes the doublet scalar $H$ with the triplet scalar $\Phi$.
To avoid this mixing, the authors of \cite{Low:2004xc} introduced
$T$-parity (similar to $R$-parity in supersymmetry) under which all
({\em but one}) new particles are odd and the SM particles are
even. Under this symmetry $H \to H$, but $\Phi \to - \Phi$, so $H^T
\Phi H$ coupling is absent. As a result, $f$ as low as 500 GeV can be
accommodated \cite{Hubisz:2005tx}. Interestingly, there exists one
new, yet $T$-even, state in this scenario, the so-called `top
partner', which cancels the standard top induced quadratic divergence
to the Higgs mass.
\begin{wrapfigure}[14]{r}{0.4\textwidth} 
\epsfxsize=0.3\textwidth
\centering \rotatebox{-90}{\epsfbox{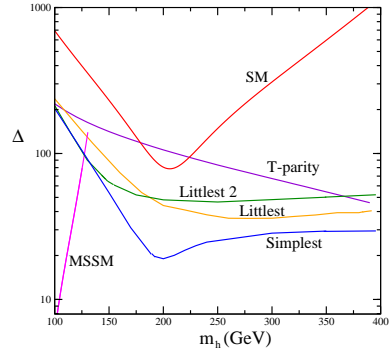}}
  \caption{\small \em  Fine-tuning in different little Higgs
    models (adapted from \cite{hep-ph/0502066}).}
\label{ft_lh} 
\end{wrapfigure}%

Remember that we set out to solve the little hierarchy problem and,
apparently, we settled that by acquiring an extra suppression factor
of $16\pi^2$. But could we actually reduce the fine-tuning in
realistic little Higgs models? Very importantly, a sizable tuning
among various contributions to the Higgs quartic coupling is necessary
to keep the Higgs mass small. Fine-tuning is relatively small when the
Higgs mass is rather high, but this option is at odds with the
requirement of EWPT. This underlines the tension between naturalness
and EWPT. In fact, fine-tuning is $\leq 1\%$ in the phenomenologically
acceptable region of the parameter space, and the general conclusion
is that little Higgs models are less natural than MSSM
\cite{hep-ph/0502066} -- see Fig.~\ref{ft_lh}.

\subsection{Collider signals of little Higgs models}
\noindent{\bf New gauge bosons:}~ In the littlest Higgs model, about
30000 $Z_H$ can be produced annually at the LHC with 100${\rm
  fb}^{-1}$ luminosity. They would decay into the SM fermions ($Z_H
\to f \bar f$), or into the SM gauge bosons ($Z_H \to W^+ W^-$, $W_H
\to W Z$, or into the Higgs and SM gauge boson ($Z_H \to Z h$). The
branching ratios would follow a definite pattern, which would serve as
the `smoking gun signals' \cite{Han:2003wu,Burdman:2002ns}.

\noindent{\bf New fermions:}~ Colored vector-like $T$ quark appears in
almost all little Higgs models. It may be produced singly by $bW \to
T$ at the LHC. Typically, $\Gamma (T \to th) \approx \Gamma (T \to tZ)
\approx \frac{1}{2} \Gamma (T \to bW)$. These branching ratio relations
would constitute a characteristic signature for $T$ quark discovery
\cite{Han:2003wu,Perelstein:2003wd}.

\noindent{\bf New scalars:}~ The presence of a doubly charged scalar
$\phi^{++}$, as a component of a complex triplet scalar, is a hallmark
signature of a large class of little Higgs models. Its decay into
like-sign dileptons ($\phi^{++} \to \ell^+ \ell^+$) would lead to an
unmistakable signal with a separable SM background \cite{Han:2003wu}.

\section{Composite Higgs}
\subsection{Basic ideas}
The composite Higgs models emerged as an improved realization of the
little Higgs scenarios both in terms of UV completion and the
naturalness consideration. In the composite picture the Higgs is some
kind of a composite bound state emerging from a strongly interacting
conformal sector \cite{hep-ph/0412089} (for a review, see
\cite{Contino:2010rs,arXiv:1109.1180} and references therein). It is a
pseudo-Goldstone boson which results when a global group $G$ of a
strongly coupled sector breaks to $H$ at a scale $f (>v)$. The coset
$G/H$ contains the Higgs.  We know that AdS/CFT correspondence allows
us to relate a strongly coupled 4d theory to a weakly coupled 5d AdS
theory. Using this correspondence, while on the CFT side the Higgs can
be viewed as a pseudo-Goldstone of some strongly coupled dynamics, on
the AdS side, in what is called the {\em Gauge-Higgs unification}
scenario \cite{Serone:2009kf}, the same Higgs can be interpreted as
the 5th component of a gauge field ($A_5$) propagating in the warped
extra dimension,

It is also called a {\em holographic Higgs} \cite{Contino:2003ve}. The
holographic 5d to 4d translation involves the presence of two sectors
- weak and strong. The weakly interacting sector containing elementary
objects (the SM gauge bosons and some fermions) is located at the
$y=0$ (Planck) brane, and the strongly interacting CFT sector at the
$y=L$ (${\rm TeV}^{-1}$) brane. The latter sector contains the TeV
bound states at a scale $\sim 1/L$, and the Higgs is one such bound
state. But to have a little hierarchy between $m_h$ and $1/L$, we
require the Higgs to be a Goldstone resulting from some $G \to H$
breaking in the CFT sector.  More precisely, the Higgs is a
pseudo-Goldstone as the couplings of the SM gauge and matter fields
with the CFT sector {\em explicitly} break $G$.

A very satisfactory feature of composite Higgs is that a non-linearly
realized global symmetry of the CFT sector protects its mass and
guarantees the absence of quadratic divergence at all order.  The
finiteness of the Higgs mass can be understood as follows: the Higgs
is at the TeV brane, the scalar that breaks the bulk gauge symmetry
lives at the Planck brane. The Higgs mass is generated by radiative
corrections with loops involving bulk KK gauge fields which propagate
from one brane to another. This mediation mechanism involves a
transmission of information from the Planck to TeV brane, which makes
it a {\em non-local} effect, and hence the potential (and, therefore,
the Higgs mass) so generated is calculable and finite.  This is a big
advantage over the conventional little Higgs construction which
suffers from quadratic cutoff sensitivity at two loop level.

Also, the global symmetry that protects the Higgs mass is a symmetry
of the strong CFT sector. Therefore, one expects to see a set of new
electroweak resonances which should appear as {\em complete}
multiplets of the global group. For example, in the ${\rm SO(5)}/{\rm
  SO(4)}$ model, additional fermionic states besides the SM fermions
are required to fill the spinorial representation ${\underline{4}}$ of
SO(5).  The spectrum of new particles can therefore reveal the nature
of the global symmetry, and is certainly richer than that of the
conventional little Higgs models.

\subsection{Collider tests} 
\noindent $(i)$~ A generic prediction of composite Higgs is that its
gauge and the Yukawa couplings are reduced from their SM values
\cite{arXiv:1003.3251}. It can be parametrized as ($\xi \equiv
\displaystyle\frac{v^2}{f^2}$)
$$
g_{hff} = g_{hff}^{\rm SM} \left(1-C_f \xi\right), ~~~
g_{hVV} = g_{hVV}^{\rm SM} \left(1-C_V \xi\right) \, .
$$ where $\xi \sim 0.2$ is small enough to keep the contribution of
the new resonances to the oblique parameters under control. Here $C_f$
and $C_V$ are numbers which depend on the choices of the groups $G$
and $H$. The question is, however, whether the Higgs production cross
section times its branching ratios in different channels can be
measured with an accuracy of, say, (10-20)\% or better? We would
perhaps need to go to super-LHC or better to ILC to confirm or rule
out compositeness in a definitive way.

\noindent $(ii)$~ Since the gauge coupling of the Higgs is smaller
than $g$, there will be incomplete cancellation of divergence in the
gauge boson scattering amplitude. As a result,
$$
A(V_LV_L \to V_LV_L) \sim s/f^2 \, . 
$$ Therefore, one hopes to see excess events in $V_LV_L \to V_LV_L$
channels. Again, this discussion is not perhaps experimentally relevant
before we reach 14 TeV, maybe not before the super-LHC stage!

\noindent $(iii)$~ The composite Higgs models usually contain heavy
colored fermions of exotic charge, e.g. electric charge $5/3$,
although this is a model dependent statement. Their production and
decay may proceed as follows:
$$q\bar q, gg \to q^*_{5/3} \bar{q}^*_{5/3} \to W^+ t W^- \bar{t} \to 
W^+W^+ b W^-W^- \bar{b} \, .
$$ The decay products contain highly energetic same sign leptons, plus
6 jets two of which two are tagged $b$ jets. Detecting those bound
states would of course constitute the best test for compositeness
\cite{arXiv:0801.1679}.

\section{Higgsless scenarios}
\begin{figure*}
%\vspace{-0.1in}
\begin{minipage}[t]{0.42\textwidth}
\epsfxsize=6cm
\centering{\epsfbox
{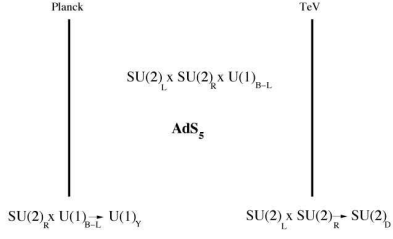}}
\end{minipage}
\hfill
\begin{minipage}[t]{0.42\textwidth}
\epsfxsize=6cm
\centering{\epsfbox
{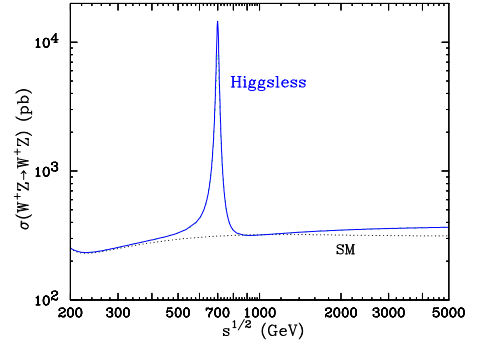}}
\end{minipage}
\vspace{5pt}
\caption[]{\small \em {(a) Left: 5d Higgsless model
    \cite{Csaki:2005vy}, (b) Right: LHC signature in Higgsless
    model (Adapted from \cite{Birkedal:2004au}).}}
\label{f_higgsless}
\end{figure*}

The idea here is to trigger electroweak symmetry breaking without
actually having a physical Higgs. This is intrinsically an extra
dimensional scenario.  The basic construction goes as follows: the
extra dimension is compactified on a circle of radius $R$ with an
orbifolding ($S^1/Z_2$). There are two fixed points: $y = 0, \pi R$.
Electroweak breaking is achieved by imposing different boundary
conditions (BC) on gauge fields at $y = 0, \pi R$ . The BCs have to be
carefully chosen such that the rank of a gauge group is lowered. The
details can be found in \cite{Csaki:2005vy}.  The extra dimension can
be flat or warped. It is difficult to control the $T$ parameter in
flat space, but in warped space one can construct a scenario which
satisfies all EWPT constraints.  Appropriate BCs are chosen to ensure
the following gauge symmetry in the bulk and in the two branes (see
Fig.~\ref{f_higgsless}a):~ Bulk:~ {${\rm SU(2)}_{\rm L} \times {\rm
    SU(2)}_{\rm R} \times {\rm U(1)}_{\rm B-L}$}, ~ $y = 0$ brane:~
${\rm SU(2)}_{\rm L} \times {\rm SU(2)}_{\rm R} \to {\rm SU(2)}_{\rm
  D}$,~ $y = \pi R$ brane:~ ${\rm SU(2)}_{\rm R} \times {\rm
  U(1)}_{\rm B-L} \to {\rm U(1)}_{\rm Y}$. Without going into the
details, for which we refer the readers to
\cite{Csaki:2005vy,Csaki:2003dt}, we mention that the $W$ and $Z$
boson masses, and the $(S,T)$ parameters can be nicely fit in a warped
scenario.

We highlight here two features which deserve attention. 

\noindent ($i$)~ {\bf Tension between unitarity and EWPT:}~ Recall
that without a Higgs, unitarity violation would set in the SM at
around a TeV. What is expected in the Higgsless scenario?  Here, the
exchange of KK states would retard the energy growth of the
$W_L$-$W_L$ scattering amplitude, {\em postponing} the violation of
unitarity in a {\em calculable} way beyond a TeV.  More specifically,
$\Lambda \sim 3\pi^4 M_W^2/(g^2 M_W^{(1)}) \sim 4$ TeV for $M_W^{(1)}
\sim 1$ TeV. If we want to postpone the onset of unitarity violation
even further, we have to decrease the $W^{(1)}$ mass. But this, in
turn, increases the $T$ parameter, implying a tension between
unitarity and EWPT \cite{tension_higgsless}.

\noindent ($ii$)~ {\bf LHC signature:}~ We deal with a specific
signature here \cite{Birkedal:2004au}. Consider the scattering channel
$WZ \stackrel{W^{(1)}}{\longrightarrow} WZ$.  If $M_W^{(1)} \approx
700$ GeV, it turns out that $g_{WZV^1} \leq \frac{g_{WWZ}
  M_Z^2}{\sqrt{3} M_W^{(1)} M_W} \sim 0.04$.  We then expect to see
sharp resonance due to $s$ channel mediation, with a striking feature
of narrow width -- see Fig.~\ref{f_higgsless}b.

%%%%%%%%%%%%%%%%%%%%%%%%%%%%%%%%%%%%%%%%%%%%%%%%%%%%%%%%%%%%%%%%%%%%
%%%%%%%%%%%%%%%%%%%%%%%%%%%%%%%%%%%%%%%%%%%%%%%%%%%%%%%%%%%%%%%%%%%%%%%%%

\section{Comparing Little Higgs/Composite Higgs/Higgsless scenarios}
\noindent $(i)$~ {\bf Little Higgs vs Composite:}~ Little Higgs models
were introduced to solve the little hierarchy problem, but these
models, as we saw before, are still quite fine-tuned. Moreover, the
Higgs mass has a quadratic cutoff sensitivity at two-loop level. The
composite Higgs does much better in both these aspects. It can have a
proper UV completion all the way to $M_{\rm Pl}$ and the Higgs mass is
finite at all order due to non-locality.

At an observational level, the composite Higgs model contains KK gluon
(as it is dual to a 5d gauge theory), while a conventional little
Higgs model does not have a KK gluon.

Fine-tuning in a composite model boils down to ensuring that $\xi =
v^2/f^2 \sim 0.2$. The new resonances in composite models weigh around
$g_\rho f$, where $g_\rho$ could be as large as $4\pi$. So the new
resonances are heavy enough and their effects on EWPT normally die
out. But there is a subtle point here which differentiates composite
Higgs from the conventional little Higgs.  The composite Higgs
couplings to the gauge bosons are different from the corresponding SM
couplings. These couplings pick up a factor $\xi$ after the Higgs
kinetic term is canonically normalized. As a result, the smooth
cancellation of log divergence between the Higgs and gauge boson
contributions to the $S$ and $T$ parameters does not hold any more,
yielding an IR contribution $\sim$ $\xi \ln(m_\rho/m_h)$ to the EWPT
parameters \cite{Barbieri:2007bh}. This contribution is numerically
very important, but a value of $\xi$ fine-tuned to approximately $0.2$
keeps the EWPT constraints under control.  In conventional little
Higgs models, the new heavy states weigh around $g f$, where $g$ is
the SM gauge coupling, and consequently the new resonances are not
heavy enough. Therefore the little Higgs resonances pose a threat to
overshoot the EWPT constraints.

\noindent $(ii)$~ {\bf Composite vs Higgsless/Technicolor:}~ In the
Technicolor model, QCD-like strong dynamics breaks electroweak
symmetry {\em directly}. The 5d Higgsless model can be seen as dual to
the `Walking Technicolor'.  In a composite Higgs model, the strong
sector does {\em not directly} break electroweak symmetry, but just
delivers a composite pseudo-Goldstone boson, the Higgs, which gets a
potential at one-loop and triggers electroweak breaking.  This
two-stage breaking, creating the parameter $\xi$, turns out to be a
boon while facing EWPT constraints.  The $S$ parameter in a composite
Higgs model is under control and, in fact, suppressed by the factor
$\xi \sim 0.2$ compared to the value of $S$ in the Higgsless model.

\noindent $(iii)$~ {\bf Interpolation:}~ The composite Higgs scenario
interpolates between the SM and the Higgsless (or, Technicolor) models
in the two extreme limits of $\xi$ \cite{arXiv:0910.4976}:
 
\centerline{${\rm {SM}} \stackrel{0 \leftarrow \xi} {\longleftarrow} {\rm
  {Composite}} \stackrel{\xi \to 1} {\longrightarrow} {\rm
 {Higgsless/TC}}$}

\section{Conclusions and Outlook}
\noindent {\bf 1.}~ 
All the BSM models we have considered are based on {\em
  calculability}.  $M_Z$ can be expressed in terms of some high scale
parameters $a_i$, i.e. $M_Z = \Lambda_{\rm NP} f(a_i)$, where $f(a_i)$
are calculable functions of physical parameters.  The new physics
scales originate from different dynamics in different cases:
$\Lambda_{\rm SUSY} \sim M_S$ (the supersymmetry breaking scale);
$\Lambda_{\rm LH} \sim f \sim F$ (the vev of $G \to H$ breaking);
$\Lambda_{\rm Extra-D} \sim R^{-1}$ (the inverse radius of
compactification).

\noindent {\bf 2.}~ In supersymmetry the cancellation of quadratic
divergence takes place between a particle loop and a sparticle
loop. Since a particle and a sparticle differ in spin, the $S,T,U$
parameters and the $Zb\bar{b}$ vertex correction can be kept under
control, since new physics appear through loops. In the little Higgs
scenario the cancellation occurs between loops with the same spin
states. Such states can mix among themselves, leading to dangerous
tree level contributions to the oblique parameters.  This is the
reason why a decoupling theory like supersymmetry is comfortable with
EWPT, while a Technicolor-like non-decoupling theory faces a stiff
confrontation.

\noindent {\bf 3.}~ We have a three-fold goal while building BSM
physics: $(i)$ unitarize the theory, $(ii)$ successfully confront EWPT,
and $(iii)$ maintain as much naturalness as possible. The tension
arises as `naturalness' demands the spectrum to be compressed, while
`EWPT compatibility' pushes the new states away from the SM states.

\noindent {\bf 4.}~ Supersymmetric theories are getting increasingly
fine-tuned with non-observation at LHC (having said that, we must
realize that LHC direct searches do not apply for third generation
matter superfields, and so a spectrum with inverted hierarchy is
relatively less tuned).  Naturalness in MSSM improved when we added
extra singlets and additional terms in the superpotential. Although
supersymmetry solves the big hierarchy problem by stabilizing the weak
scale over many decades in energy scale, the little hierarchy problem
continues to haunt and instigate the model builders to take bold,
sometimes outrageous, steps for reducing the fine-tuning.

\noindent {\bf 5.}~ A light Higgs need not necessarily be
elementary. It can very well be a composite object. Also, a narrow
width of Higgs does not necessarily attest its elementarity. A light
composite Higgs can very well have a narrow width. Just finding the
Higgs would not settle this issue. We need to measure the Higgs
couplings very precisely to know whether it is elementary or
composite.

\noindent {\bf 6.}~ LHC is a `win-win' discovery machine.  If we find
the Higgs, it will be a great discovery. But if we see only the Higgs
and nothing else, we will definitely be disappointed as many of our
questions would be left unanswered \cite{arXiv:1110.3805}. If, on the
other hand, LHC confirms that there is no Higgs, it will be no less a
discovery \cite{cerndg}.  If the Higgs is not there, the new
resonances which would restore unitarity in gauge boson scattering
should show up, with a prior hint of excess events in $V_LV_L \to
V_LV_L$ scattering.  In that case we absolutely need the super-LHC,
and eventually the ILC, to confidently establish the nature of the new
resonances.

\noindent {\bf 7.}~ The excluded region in BSM parameter space is
growing fast as LHC accumulate more and more data \cite{henri}.  By
the time we meet in the next Lepton Photon Conference in 2013, many of
the possibilities discussed here may not perhaps be heard again!  But
who can rule out the possibility that completely new theoretical ideas
inspired by yet unseen unexpected observations during the next two
years would form the cornerstone for building the physics of the TeV
scale?

\noindent{\em Note added:}~ On 13 December 2011 at CERN, the ATLAS and
CMS experiments presented their update on the Higgs searches. The SM
Higgs mass is now allowed in a narrower window: $115 < m_h^{\rm SM} <
127$ GeV at 95\% C.L. with a mild excess around 125 GeV.  The main
ethos of this talk remains unaffected by this observation. The only
observation we can make at this stage is that if the Higgs mass is
later confirmed to be around 125 GeV, some of the supersymmetric
models we have discussed would require more tuning than before.

\noindent{\bf Acknowledgements:}~ I thank the organizers of Lepton
Photon 2011, especially Naba Mandal and Rohini Godbole, for arranging
such a stimulating conference and for inviting me to give this talk. I
acknowledge hospitality of the CERN PH/TH division during the writing
of this draft. I also thank Emilian Dudas, Christophe Grojean, Palash
B. Pal, and Amitava Raychaudhuri for discussions and clarifications on
several issues.


\begin{thebibliography}{99}
\small

%\cite{arXiv:0910.5095}
\bibitem{arXiv:0910.5095}
  G.~Bhattacharyya,
  %``A Pedagogical Review of Electroweak Symmetry Breaking Scenarios,''
  Rept.\ Prog.\ Phys.\ \ {\bf 74} (2011) 026201
  [arXiv:0910.5095 [hep-ph]], and references therein. 
  %%CITATION = RPPHA,74,026201;%%

%\cite{Giudice:2008bi}
\bibitem{Giudice:2008bi}
  G.~F.~Giudice,
  %``Naturally Speaking: The Naturalness Criterion and Physics at the LHC,''
  arXiv:0801.2562 [hep-ph].
  %%CITATION = ARXIV:0801.2562;%%


%\cite{Contino:2010rs}
\bibitem{Contino:2010rs}
  R.~Contino,
  %``Tasi 2009 lectures: The Higgs as a Composite Nambu-Goldstone Boson,''
  arXiv:1005.4269 [hep-ph].
  %%CITATION = ARXIV:1005.4269;%%

%\cite{arXiv:0910.4976}
\bibitem{arXiv:0910.4976}
  C.~Grojean,
  %``New theories for the Fermi scale,''
  PoSEPS\ {\bf -HEP2009} (2009) 008
  [arXiv:0910.4976 [hep-ph]].
  %%CITATION = POSCI,EPS-HEP2009,008;%%


\bibitem{susy-books} See the text books on supersymmetry, \\
  R.N. Mohapatra, ``Unification and Supersymmetry: The Frontiers of
  quark-lepton physics,'' Springer-Verlag, NY 1992; M.~Drees, R.~Godbole
  and P.~Roy, ``Theory and phenomenology of sparticles: An account of
  four-dimensional N=1 supersymmetry in high energy physics,'' World
  Scientific (2004); H.~Baer and X.~Tata, ``Weak scale supersymmetry: From
  superfields to scattering events,'' Cambridge, UK: Univ. Pr. (2006). See
  also, R.K. Kaul, ``Supersymmetry and supergravity:'' in {\em Gravitation,
    gauge theories and the early universe}, 487-522, Ed. B.R. Iyer, Kluwer
  Academic Publishers (1989).


%\cite{CERN-TH-4825/87}
\bibitem{CERN-TH-4825/87}
  R.~Barbieri and G.~F.~Giudice,
  %``Upper Bounds on Supersymmetric Particle Masses,''
  Nucl.\ Phys.\ B\ {\bf 306} (1988) 63.
  %%CITATION = NUPHA,B306,63;%%

%\cite{arXiv:1101.2195}
\bibitem{arXiv:1101.2195}
  A.~Strumia,
  %``The Fine-tuning price of the early LHC,''
  JHEP\ {\bf 1104} (2011) 073
  [arXiv:1101.2195 [hep-ph]].
  %%CITATION = JHEPA,1104,073;%%

%\cite{arXiv:1107.2472}
\bibitem{arXiv:1107.2472}
  U.~Ellwanger, G.~Espitalier-Noel and C.~Hugonie,
  %``Naturalness and Fine Tuning in the NMSSM: Implications of Early LHC Results,''
  arXiv:1107.2472 [hep-ph].
  %%CITATION = ARXIV:1107.2472;%%

%\cite{arXiv:1101.4664}
\bibitem{arXiv:1101.4664}
  S.~Cassel, D.~M.~Ghilencea, S.~Kraml, A.~Lessa and G.~G.~Ross,
  %``Fine-tuning implications for complementary dark matter and LHC SUSY searches,''
  JHEP\ {\bf 1105} (2011) 120
  [arXiv:1101.4664 [hep-ph]].
  %%CITATION = JHEPA,1105,120;%%


\bibitem{nmssm} For reviews on NMSSM, see
  U.~Ellwanger, C.~Hugonie and A.~M.~Teixeira, 
%``The Next-to-Minimal Supersymmetric Standard Model,'' 
Phys.\ Rept.\ {\bf 496} (2010) 1
  [arXiv:0910.1785 [hep-ph]]; M.~Maniatis, 
%``The Next-to-Minimal
%  Supersymmetric extension of the Standard Model reviewed,''
  Int.\ J.\ Mod.\ Phys.\ A {\bf 25} (2010) 3505 [arXiv:0906.0777
    [hep-ph]].
  

%\cite{Drees:1988fc}
\bibitem{Drees:1988fc}
  M.~Drees,
%  ``Supersymmetric Models with Extended Higgs Sector,''
  Int.\ J.\ Mod.\ Phys.\  A {\bf 4} (1989) 3635.
  %%CITATION = IMPAE,A4,3635;%%

%\cite{arXiv:1108.1284}
\bibitem{arXiv:1108.1284}
  G.~G.~Ross and K.~Schmidt-Hoberg,
  %``The fine-tuning and phenomenology of the generalised NMSSM,''
  arXiv:1108.1284 [hep-ph].
  %%CITATION = ARXIV:1108.1284;%%


%\cite{hep-ph/9506359}
\bibitem{hep-ph/9506359}
  S.~A.~Abel, S.~Sarkar and P.~L.~White,
  %``On the cosmological domain wall problem for the minimally extended supersymmetric standard model,''
  Nucl.\ Phys.\ B\ {\bf 454} (1995) 663
  [hep-ph/9506359].
  %%CITATION = NUPHA,B454,663;%%


%\cite{arXiv:1004.1271}
\bibitem{arXiv:1004.1271}
  P.~Lodone,
  %``Naturalness bounds in extensions of the MSSM without a light Higgs boson,''
  JHEP\ {\bf 1005} (2010) 068
  [arXiv:1004.1271 [hep-ph]].
  %%CITATION = JHEPA,1005,068;%%

%\cite{hep-ph/9611243}
\bibitem{hep-ph/9611243}
  G.~Bhattacharyya and A.~Romanino,
  %``Naturalness constraints on gauge mediated supersymmetry breaking models,''
  Phys.\ Rev.\ D\ {\bf 55} (1997) 7015
  [hep-ph/9611243].
  %%CITATION = PHRVA,D55,7015;%%


\bibitem{lhreviews}
M.~Schmaltz and D.~Tucker-Smith,
%  ``Little Higgs review,'' 
  Ann.\ Rev.\ Nucl.\ Part.\ Sci.\  {\bf 55} (2005) 229
  [arXiv:hep-ph/0502182]; 
%\cite{Perelstein:2005ka}
%\bibitem{Perelstein:2005ka}
  M.~Perelstein,
%  ``Little Higgs models and their phenomenology,''
  Prog.\ Part.\ Nucl.\ Phys.\  {\bf 58} (2007) 247
  [arXiv:hep-ph/0512128]; 
%\cite{Chen:2006dy}
%\bibitem{Chen:2006dy}
  M.~C.~Chen,
%  ``Models of little Higgs and electroweak precision tests,''
  Mod.\ Phys.\ Lett.\  A {\bf 21} (2006) 621
  [arXiv:hep-ph/0601126].
  %%CITATION = MPLAE,A21,621;%%


%\cite{Cheng:2007bu}
\bibitem{Cheng:2007bu}
  H.~C.~Cheng,
 % ``Little Higgs, Non-standard Higgs, No Higgs and All That,''
  arXiv:0710.3407 [hep-ph].
  %%CITATION = ARXIV:0710.3407;%%


%\cite{Kaul:2008cv}
\bibitem{Kaul:2008cv}
  R.~K.~Kaul,
 % ``Naturalness and Electro-weak Symmetry Breaking,''
  arXiv:0803.0381 [hep-ph].
  %%CITATION = ARXIV:0803.0381;%%


\bibitem{large_f_ewpt}
%\cite{Csaki:2002qg}
%\bibitem{Csaki:2002qg}
  C.~Csaki, J.~Hubisz, G.~D.~Kribs, P.~Meade and J.~Terning,
%  ``Big corrections from a little Higgs,''
  Phys.\ Rev.\  D {\bf 67} (2003) 115002
  [arXiv:hep-ph/0211124]; 
  %%CITATION = PHRVA,D67,115002;%%
%\cite{Hewett:2002px}
%\bibitem{Hewett:2002px}
  J.~L.~Hewett, F.~J.~Petriello and T.~G.~Rizzo,
%  ``Constraining the littlest Higgs. ((U)),''
  JHEP {\bf 0310} (2003) 062
  [arXiv:hep-ph/0211218]; 
  %%CITATION = JHEPA,0310,062;%%
%\cite{Chen:2003fm}
%\bibitem{Chen:2003fm}
  M.~C.~Chen and S.~Dawson,
%  ``One-loop radiative corrections to the rho parameter in the littlest  Higgs
  model,''
  Phys.\ Rev.\  D {\bf 70} (2004) 015003
  [arXiv:hep-ph/0311032].
  %%CITATION = PHRVA,D70,015003;%%


%\cite{Low:2004xc}
\bibitem{Low:2004xc}
  I.~Low,
%  ``T parity and the littlest Higgs,''
  JHEP {\bf 0410} (2004) 067
  [arXiv:hep-ph/0409025]; 
  %%CITATION = JHEPA,0410,067;%%
%\cite{Cheng:2004yc}
%\bibitem{Cheng:2004yc}
  H.~C.~Cheng and I.~Low,
%  ``Little hierarchy, little Higgses, and a little symmetry,''
  JHEP {\bf 0408} (2004) 061
  [arXiv:hep-ph/0405243].
  %%CITATION = JHEPA,0408,061;%%


%\cite{Hubisz:2005tx}
\bibitem{Hubisz:2005tx}
  J.~Hubisz, P.~Meade, A.~Noble and M.~Perelstein,
%  ``Electroweak precision constraints on the littlest Higgs model with T parity,''
  JHEP {\bf 0601} (2006) 135
  [arXiv:hep-ph/0506042].
  %%CITATION = JHEPA,0601,135;%%


%\cite{hep-ph/0502066}
\bibitem{hep-ph/0502066}
  J.~A.~Casas, J.~R.~Espinosa and I.~Hidalgo,
  %``Implications for new physics from fine-tuning arguments. II. Little Higgs models,''
  JHEP\ {\bf 0503} (2005) 038
  [hep-ph/0502066].
  %%CITATION = JHEPA,0503,038;%%

%\cite{Han:2003wu}
\bibitem{Han:2003wu}
  T.~Han, H.~E.~Logan, B.~McElrath and L.~T.~Wang,
%  ``Phenomenology of the little Higgs model,''
  Phys.\ Rev.\  D {\bf 67} (2003) 095004
  [arXiv:hep-ph/0301040].
  %%CITATION = PHRVA,D67,095004;%%

%\cite{Burdman:2002ns}
\bibitem{Burdman:2002ns}
  G.~Burdman, M.~Perelstein and A.~Pierce,
%  ``Collider tests of the little Higgs model,''
  Phys.\ Rev.\ Lett.\  {\bf 90} (2003) 241802
  [Erratum-ibid.\  {\bf 92} (2004) 049903]
  [arXiv:hep-ph/0212228].
  %%CITATION = PRLTA,90,241802;%%

%\cite{Perelstein:2003wd}
\bibitem{Perelstein:2003wd}
  M.~Perelstein, M.~E.~Peskin and A.~Pierce,
 % ``Top quarks and electroweak symmetry breaking in little Higgs models,''
  Phys.\ Rev.\  D {\bf 69} (2004) 075002
  [arXiv:hep-ph/0310039].
  %%CITATION = PHRVA,D69,075002;%%

%\cite{hep-ph/0412089}
\bibitem{hep-ph/0412089}
  K.~Agashe, R.~Contino and A.~Pomarol,
  %``The Minimal composite Higgs model,''
  Nucl.\ Phys.\ B\ {\bf 719} (2005) 165
  [hep-ph/0412089].
  %%CITATION = NUPHA,B719,165;%%

%\cite{arXiv:1109.1180}
\bibitem{arXiv:1109.1180}
  S.~Rychkov,
  %``EWSB Theory on the Eve of Higgs Boson Exclusion/Discovery,''
  arXiv:1109.1180 [hep-ph].
  %%CITATION = ARXIV:1109.1180;%%


%\cite{Serone:2009kf}
\bibitem{Serone:2009kf}
  M.~Serone,
%  ``Holographic Methods and Gauge-Higgs Unification in Flat Extra Dimensions,''
  New J.\ Phys.\  {\bf 12} (2010) 075013
  [arXiv:0909.5619 [hep-ph]].
  %%CITATION = NJOPF,12,075013;%%

%\cite{Contino:2003ve}
\bibitem{Contino:2003ve}
  R.~Contino, Y.~Nomura and A.~Pomarol,
%  ``Higgs as a holographic pseudo-Goldstone boson,''
  Nucl.\ Phys.\  B {\bf 671} (2003) 148
  [arXiv:hep-ph/0306259].
  %%CITATION = NUPHA,B671,148;%%



%\cite{arXiv:1003.3251}
\bibitem{arXiv:1003.3251}
  J.~R.~Espinosa, C.~Grojean and M.~Muhlleitner,
  %``Composite Higgs Search at the LHC,''
  JHEP\ {\bf 1005} (2010) 065
  [arXiv:1003.3251 [hep-ph]].
  %%CITATION = JHEPA,1005,065;%%


%\cite{arXiv:0801.1679}
\bibitem{arXiv:0801.1679}
  R.~Contino and G.~Servant,
  %``Discovering the top partners at the LHC using same-sign dilepton final states,''
  JHEP\ {\bf 0806} (2008) 026
  [arXiv:0801.1679 [hep-ph]].
  %%CITATION = JHEPA,0806,026;%%



%\cite{Csaki:2005vy}
\bibitem{Csaki:2005vy}
  For a review on Higgsless models, see C.~Csaki, J.~Hubisz and P.~Meade,
%  ``Electroweak symmetry breaking from extra dimensions,''
  arXiv:hep-ph/0510275 (TASI lectures).
  %%CITATION = HEP-PH/0510275;%%

%\cite{Csaki:2003dt}
\bibitem{Csaki:2003dt}
  C.~Csaki, C.~Grojean, H.~Murayama, L.~Pilo and J.~Terning,
%  ``Gauge theories on an interval: Unitarity without a Higgs,''
  Phys.\ Rev.\  D {\bf 69} (2004) 055006
  [arXiv:hep-ph/0305237].
  %%CITATION = PHRVA,D69,055006;%%



\bibitem{tension_higgsless}
  R.~Barbieri, A.~Pomarol and R.~Rattazzi,
%  ``Weakly coupled Higgsless theories and precision electroweak tests,''
  Phys.\ Lett.\  B {\bf 591} (2004) 141
  [arXiv:hep-ph/0310285]; 
  %%CITATION = PHLTA,B591,141;%%
G.~Cacciapaglia, C.~Csaki, C.~Grojean and J.~Terning,
%  ``Oblique corrections from Higgsless models in warped space,''
  Phys.\ Rev.\  D {\bf 70} (2004) 075014
  [arXiv:hep-ph/0401160]; 
%\cite{Davoudiasl:2003me}
%\bibitem{Davoudiasl:2003me}
  H.~Davoudiasl, J.~L.~Hewett, B.~Lillie and T.~G.~Rizzo,
%  ``Higgsless electroweak symmetry breaking in warped backgrounds:  Constraints
%  and signatures,''
  Phys.\ Rev.\  D {\bf 70} (2004) 015006
  [arXiv:hep-ph/0312193].
  %%CITATION = PHRVA,D70,015006;%%


%\cite{Birkedal:2004au}
\bibitem{Birkedal:2004au}
  A.~Birkedal, K.~Matchev and M.~Perelstein,
%  ``Collider phenomenology of the Higgsless models,''
  Phys.\ Rev.\ Lett.\  {\bf 94} (2005) 191803
  [arXiv:hep-ph/0412278].
  %%CITATION = PRLTA,94,191803;%%


%\cite{Barbieri:2007bh}
\bibitem{Barbieri:2007bh}
  R.~Barbieri, B.~Bellazzini, V.~S.~Rychkov and A.~Varagnolo,
  %``The Higgs boson from an extended symmetry,''
  Phys.\ Rev.\ D {\bf 76} (2007) 115008
  [arXiv:0706.0432 [hep-ph]]; 
  %%CITATION = ARXIV:0706.0432;%%
%\cite{Espinosa:2010vn}
%\bibitem{Espinosa:2010vn}
  J.~R.~Espinosa, C.~Grojean and M.~Muhlleitner,
  %``Composite Higgs Search at the LHC,''
  JHEP {\bf 1005} (2010) 065
  [arXiv:1003.3251 [hep-ph]].
  %%CITATION = ARXIV:1003.3251;%%


%\cite{arXiv:1110.3805}
\bibitem{arXiv:1110.3805}
  M.~E.~Peskin,
  ``Summary of Lepton Photon 2011,''
  arXiv:1110.3805 [hep-ph] (In these proceedings).
  %%CITATION = ARXIV:1110.3805;%%


\bibitem{cerndg} R. Heuer, Talk in Lepton-Photon 2011, titled `Whither
  Colliders After LHC ?' (In these proceedings).

\bibitem{henri} H. Bachacou, Talk in Lepton-Photon 2011, titled `BSM
  Results from LHC' (In these proceedings).


\end{thebibliography}
\end{document}